\documentclass[prd,nofootinbib,twocolumn,showpacs,preprintnumbers,amsmath,amssymb]{revtex4}
\usepackage{graphicx}
\usepackage{dcolumn}
\usepackage{bm}
\usepackage{wasysym}
\usepackage{float} 
\usepackage{csquotes}

\newcommand{\apjs}{ApJS}

\newcommand{\mnras}{MNRAS}

\newcommand{\jcap}{JCAP}

\usepackage{euscript}

\DeclareMathOperator{\arcosh}{arcosh}
\DeclareMathOperator{\divi}{div}

\begin{document}

\preprint{APS/123-QED}

\title{How empty are the voids in the large-scale structure of the Universe?}

\author{A. N. Baushev}
\email{baushev@gmail.com}

\affiliation{Bogoliubov Laboratory of Theoretical Physics, Joint Institute for Nuclear Research,
141980 Dubna, Moscow Region, Russia\\ The department of Physics, Helsinki University, Helsinki, Finland}

\date{\today}

\begin{abstract}
We find an analytical solution for the minimal matter density of a void, its central density. The solution suggests that voids are not as empty as they seem: most voids have a central underdensity $\Delta_c \sim -50\%$ (which means that the matter density at their centers is only a factor of two lower than in the Universe on average). Small voids (with radius $R_0\simeq 5-10$~{Mpc}) can be significantly emptier, but the number of voids decreases rapidly with increasing $|\Delta_c|$ above $50\%$, and voids with $\Delta_c < -80\%$ are practically absent. Large voids ($R_0\ge 40$~{Mpc}) always have $|\Delta_c| < 50\%$. We find a clear discrepancy between the analytical estimate of the central density of voids and the results of N-body simulations, the nature of which is not entirely clear.
\end{abstract}

\pacs{95.10.-a; 98.80.Es; 98.35.Ce; 98.56.-p; 98.62.Ck; 98.65.Cw}

\maketitle

\section{Introduction}
Voids are vast ($\sim 10-100$~{Mpc}) regions of the large-scale structure of the Universe that contain no galaxy clusters and almost no bright galaxies. Voids are ellipsoidal and occupy a significant fraction of the Universe's volume. The ratio between the matter density inside voids and the average density of the Universe remains unclear. On the one hand, the density of bright galaxies is tens of times lower inside voids than on average in the Universe, which implies that voids are indeed quite empty. On the other hand, in voids one can easily observe only the stellar component of sufficiently bright galaxies. There are reasons to believe that structure formation in voids is suppressed \cite{26}, and a significant amount of matter there forms a massive diffuse component. Voids may contain an almost arbitrary amount of dark matter (hereafter DM), both as a homogeneous component and as dark halos. Baryonic matter in the form of dwarf galaxies, hot gas, ultradiffuse galaxies (UDGs), etc. is also barely detectable in voids. Therefore, the actual emptiness of voids is difficult to measure. 

N-body simulations support the view that voids are highly underdense: they suggest that the matter density at the void center is almost ten times lower than the average density of the Universe (see section~\ref{27section4} for details). However, simulations may suffer from numerical effects, and it would be very useful to verify them by analytical calculations. Analytical investigations of void formation have a long history; here we mention only a few remarkable works from the extensive literature: \cite{bertschinger1985, blumenthal1992, weygaert1993}. As a rule, calculations were carried out under the assumption of spherical symmetry; in addition, in early works the cosmological constant was ignored for obvious reasons. Spherical symmetry allows one to find a solution for any initial conditions. However, the result depends on the shape of the initial perturbation, which is unknown. In addition, spherical symmetry is obviously violated when voids enter the nonlinear regime and begin to \enquote{collide} with each other, forming a flat wall between them.

In this paper, we address a much simpler problem --- estimating the minimal density in a void, i.e., the density at its center. As we show, the density profile of voids is very flat at the center, and thus the region where the density is close to the minimal value may occupy a significant fraction of the void volume. Unlike the full density profile, the minimal void density can be found analytically, and it depends only on the amplitude of the initial perturbation, which can be determined assuming that the primordial perturbations were Gaussian. We use the analytical method proposed in~\cite{26}, correcting a minor calculation error in that work and generalizing the result to the case of a non-spherical void.

The structure of the paper is as follows: in section~\emph{Calculations} section, we calculate the central density of a void in the spherically symmetric case; in section~\emph{The case of an ellipsoidal void} we consider a non-spherical void and show that its central density should be the same as in the spherically symmetric case; section~\emph{Are the voids empty?} describes the obtained results and compares them with the results of N-body simulations. 
%In section~\emph{The reason for the discrepancy} we consider the tension between the analytical estimate that we obtained and N-body simulation results, which suggest that the voids should be substantially emptier. 
Finally, section~\emph{Observational properties of voids} considers some observable void parameters that may allow one to estimate the actual void emptiness.

\section{Calculations}
We now find the central density of a spherically symmetric void. First, we need to specify the initial conditions. We choose a redshift $z_1$ deep in the matter-dominated era of the Universe, so that we may already neglect the radiation term, but the perturbation that later evolves into the void is still small --- say, $z_1 = 10$. We will denote all quantities at this moment by the subscript \enquote{1}. Let the characteristic size of the perturbation at this moment be $\lambda_1$.

The future void contains many substructures, and therefore the actual density profile of the underdensity at $z=z_1$ is quite complex. However, the substructures have little effect on void formation, and we may eliminate them by \enquote{averaging} the velocity and density profiles (for instance, with the help of a Gaussian filter). Hereafter, we will consider this \enquote{softened} profile instead of the actual one.

We denote the scale factor of the Universe by $a(t)$, and the Hubble parameter by $H\equiv\dfrac{da}{a dt}$. The redshift is related to $a$ by the equation $z+1=a_0/a$. The subscript \enquote{0} corresponds to the present-day values of the quantities. We choose the point of minimum density at the void center as the origin of coordinates and surround it by a sphere of radius $R_1\ll \lambda_1$, which expands at the same rate as the Universe: $R=R_1 \dfrac{a}{a_1}$. Since the size of the perturbation $\lambda$ changes in the same way, the ratio $R/\lambda$ remains constant. We should emphasize that the sphere is not quite comoving with the matter: the void center expands slightly faster than the Universe on average, and therefore the matter crosses the sphere outward.

 The outer regions of the void do not influence the matter inside $R$, since a spherically symmetric shell does not create a gravitational field inside itself (Birkhoff's theorem). Thus, although a void is always surrounded by vast masses of matter, it can hardly create a strong gravitational field at the void center. As we show below, this conclusion is very likely valid even for a non-spherical void. The only significant perturbations inside $R$ may be created by other nearby elements of the cosmic web, outside the void and its walls, for instance, by other nearby elements of the cosmic web beyond the void and its walls, for instance, by the walls of a neighboring void, since these structures are distributed asymmetrically with respect to the void under consideration. However, these tidal perturbations are small on the scale $R$ (they are $\sim (R/\lambda)^3$), and we neglect them.
%%%%%%%%%%%%%%%%%%%%%%%%%%%%%%%%%%%%%%%%%%%%%%%%%%%%%%%%%%%%%%%%%%%%%%%%%%%%%%%%%%%%%%%%%
 The matter density $q_m$ inside $R$ does not depend on the coordinates (since $r=0$ is an extremum of $\rho(r)$), and the matter velocity is proportional to $\vec r$: $\vec v\propto \vec r$ (due to the spherical symmetry). Moreover, the only component that has cosmologically significant pressure ($p\sim\rho$) after $z_1$ is dark energy, and its pressure is constant everywhere. Thus, the system has no pressure gradients. 
 
 We conclude that the evolution of the Universe inside $R$ may be described by the standard Friedmann equation~\cite{tolman1934}, though its parameters may differ slightly from those of the background Universe. We denote the critical, matter, dark energy, and curvature densities of the undisturbed Universe by $\rho_c, \rho_m, \rho_\Lambda, \rho_a$, of the \enquote{universe} inside $R$ --- by $q_c, q_m, q_\Lambda, q_a$, respectively. We would like to recall that the curvature density 
 $$
 \rho_a\equiv \dfrac{3c^2}{8\pi G}\dfrac{y}{a^2},
 $$
 where $y=-1, 1, 0$ for a closed, open, and flat Friedmann universe, respectively.  We denote the scale of the 'universe' inside $R$ by $b(t)$, and the Hubble parameter by $Q\equiv\dfrac{db}{b dt}$. The critical densities are 
\begin{equation}
\rho_c=\frac{3H^2}{8\pi G}; \qquad  q_c=\frac{3Q^2}{8\pi G}\label{27a1}
\end{equation}
We assume the standard $\Lambda$CDM cosmological model, i.e., the dark energy is a non-zero cosmological constant ($\rho_\Lambda=q_\Lambda= \mathit{const}$), and the Universe is flat ($\rho_a$ (but not $q_a$!) is equal to zero). Since we consider the Universe at $z\le 10$, we neglect the radiation term, and therefore $\Omega_m+\Omega_\Lambda\equiv\dfrac{\rho_m}{\rho_c}+\dfrac{\rho_\Lambda}{\rho_c}=1$. The present-day values of the ratios are $\Omega_{m,0}=0.315$ and  $\Omega_{\Lambda,0}=0.685$, and the Hubble constant is $H_0=73$~(km/s) Mpc$^{-1}$ \cite{pdg2022}. It is convenient to introduce $h$ so that $H_0=h\cdot 100$~(km/s) Mpc$^{-1}$. Thus, $h=0.73$.  

Since the Universe is flat, $a(t)$ is defined up to an arbitrary factor (whereas $b(t)$ is the radius of the 3-space of the \enquote{universe} inside $R$, and hence it is defined uniquely). Thus, we may always choose $a_1=b_1$. The Friedmann equations for the Universe and the central region of the void may be written as
\begin{eqnarray}
H^2=\frac{8\pi G}{3}\left[\rho_\Lambda+\rho_{m,1}\left(\dfrac{a_1}{a}\right)^3\right],\label{27a2}\\
Q^2=\frac{8\pi G}{3}\left[q_\Lambda+q_{m,1}\left(\dfrac{b_1}{b}\right)^3+q_{a,1}\left(\dfrac{b_1}{b}\right)^2\right].\label{27a3}
\end{eqnarray}
The values of $\rho_\Lambda$, $q_\Lambda$, and $\rho_{m,1}$ can be easily expressed through $\Omega_{m,0}$ and $\rho_{c,0}=3H_0^2/(8\pi G)$:
\begin{equation}
\rho_{m,1}=\left(\frac{a_0}{a_1}\right)^3 \rho_{c,0} \Omega_{m,0}; \quad  \rho_\Lambda = q_\Lambda=\rho_{c,0} (1-\Omega_{m,0}).\label{27a4}
\end{equation}
At $z=z_1$ the future void is just a shallow underdensity $|\delta_1|=|\Delta\rho_m/\rho_{m,1}|\ll 1$. The density contrast at the matter-dominated stage is proportional to $a(t)$ \cite{gorbrub2}, and it is more convenient to characterize the underdensity by 
\begin{equation}
\aleph\equiv (z+1) \delta=\left(\frac{a_0}{a}\right) \delta;\quad \aleph= \left(\frac{a_0}{a_1}\right) \delta_1,\label{27a5}
\end{equation}
since $\aleph$ is constant on the matter-dominated stage, and it does not depend on the choice of $z_1$. Thus,
\begin{eqnarray}
q_{m,1}=(1-\delta_1)\left(\frac{a_0}{a_1}\right)^3 \rho_{c,0} \Omega_{m,0},\label{27a6}\\
q_{m}=(1-\delta(t))\left(\frac{a_0}{a}\right)^3 \rho_{c,0} \Omega_{m,0}\label{27a7}
\end{eqnarray}
As we have already mentioned, the sphere $R$ is not exactly comoving with the matter: the void center expands slightly faster than the Universe on average, and therefore the matter crosses the sphere outward at a speed $v(t)$. The speed of sphere $R$ with respect to the origin of coordinates at $z=z_1$ is $H_1 R_1$, while the matter velocity is $Q_1 R_1$. Thus,
\begin{equation}
Q_1 R_1= H_1 R_1 +v_1.\label{27a8}
\end{equation}
The speed can be easily calculated from the continuity consideration. The flux of matter through sphere $R$ at $z=z_1$ is (we substitute eqn.~(\ref{27a6}) for $q_{m,1}$)
\begin{equation}
\frac{dM}{dt}=-4\pi R^2_1 v_1 q_{m,1}= -4\pi R^2_1 v_1 (1-\delta_1)\left(\frac{a_0}{a_1}\right)^3 \rho_{c,0} \Omega_{m,0}.\label{27a9}
\end{equation}
On the other hand,
\begin{eqnarray}
M(t)=\frac43\pi R^3_1 \left(\frac{a}{a_1}\right)^3 q_{m}=\nonumber\\
=\frac43\pi R^3_1 (1-\delta(t))\left(\frac{a_0}{a_1}\right)^3 \rho_{c,0} \Omega_{m,0}.\label{27a10}
\end{eqnarray}
Here we substitute eqn.~(\ref{27a7}) for $q_{m}$. Only multiplier $(1-\delta(t))$ depends on time in the obtained equation. We may calculate the time derivative of it with the help of eqn.~(\ref{27a5}): since $\aleph$ is constant on the matter-dominated stage, $d\aleph=0$ and from~(\ref{27a5}) it follows that  
\begin{equation}
a(t) d\delta(t)=\delta(t) d a(t);\quad \frac{d\delta(t)}{dt}=\delta(t)\frac{da}{adt}=\delta(t) H \label{27a11}
\end{equation}
Then we obtain from~(\ref{27a10}):
\begin{equation}
\frac{dM}{dt}
=\delta(t) H(t) \frac43\pi R^3_1 \left(\frac{a_0}{a_1}\right)^3 \rho_{c,0} \Omega_{m,0}.\label{27a12}
\end{equation}
Equating equations~(\ref{27a9}) and~(\ref{27a12}) at $z=z_1$, we obtain
\begin{equation}
v_1=\frac13 \delta_1 R_1 H_1
\label{27a13}
\end{equation}
As we can see, $v_1/R_1\ll H_1$, the velocity perturbation is small with respect to the Hubble expansion. Then we may rewrite~(\ref{27a8}) as $Q_1 = (1+\delta_1/3) H_1$. Squaring it and neglecting the small $\delta^2_1$ term, we obtain $Q^2_1 = (1+2\delta_1/3) H^2_1$. Substituting here equations~(\ref{27a2}) and~(\ref{27a13}) at $z=z_1$, we get:
\begin{equation}
(1+\frac23\delta_1)(\rho_\Lambda+\rho_{m,1})=q_\Lambda+q_{m,1}+q_{a,1}.
\label{27a14}
\end{equation}
In accordance with~(\ref{27a4}) and~(\ref{27a6}), $\rho_\Lambda=q_\Lambda$, $q_{m,1}=(1-\delta_1)\rho_{m,1}$, and we may convert~(\ref{27a14}) into
\begin{equation}
q_{a,1}=\frac23 \delta_1\rho_\Lambda+\frac53 \delta_1 \rho_{m,1}.
\label{27a15}
\end{equation}
However, $\rho_\Lambda$ is $\sim 3$ orders of magnitude smaller, than $\rho_{m,1}$, and we may neglect the first term. Thus
\begin{eqnarray}
q_{a,1}=\frac53 \delta_1 \rho_{m,1}=\frac53 \delta_1 \left(\frac{a_0}{a_1}\right)^3 \rho_{c,0} \Omega_{m,0}=\nonumber\\
=\frac53 \aleph \left(\frac{a_0}{a_1}\right)^2 \rho_{c,0} \Omega_{m,0}.
\label{27a16}
\end{eqnarray}
We used~(\ref{27a5}) in the last transformation. Now we substitute equations~(\ref{27a4}), (\ref{27a6}), and~(\ref{27a16}) for $q_\Lambda$, $q_{m,1}$, and $q_{a,1}$ into~(\ref{27a3}). We reduce $b_1$ and use equation~(\ref{27a1}) in the form of $8\pi G\rho_{c,0}=3H_0^2$:
\begin{eqnarray}
Q^2=H^2_0\left[\Omega_{\Lambda,0}+\Omega_{m,0} \left(\dfrac{a_0}{b}\right)^3
+\frac53 \aleph\: \Omega_{m,0}\left(\dfrac{a_0}{b}\right)^2\right].\label{27a17}
\end{eqnarray}
We also neglected the small quantity $\delta_1$ ($(1-\delta_1)\simeq 1$) in equation~(\ref{27a6}) for $q_{m,1}$. This equation\footnote{In Ref.~\cite{26}, a minor mathematical error was made: the numerical coefficient before the numerical coefficient of the last term under the square root in expression~(\ref{27a17}) for $Q$ was $3$ instead of the correct value $5/3$, as adopted here.} for $Q(t)$ determines the evolution of the region inside $R$. In order to finish our calculation, we use a well-known trick~\cite{24, 25}: the age of the Universe, as well as that of the \enquote{universe} inside $R$, may be found from the Hubble parameters:  
\begin{equation}
t_U=\int dt=\int_0^{a_0}\frac{da}{aH}, \quad t_R=\int_0^{b_0}\frac{db}{bQ},
\label{27a18}
\end{equation}
where $H(a)$ and $Q(b)$ are given by equations~(\ref{27a2}) and (\ref{27a17}), respectively. Obviously, $t_U$ must be equal to $t_R$. 

It is essential that $b_0\ne a_0$: the void center expands much more strongly than the Universe on average. Let us denote $k=b_0/a_0$. Clearly, the central density of a void is $q_{m,0}=\rho_{m,0}/k^3$. As is customary, we characterize voids by their present-day central underdensity:
\begin{equation}
\Delta_c\equiv \frac{q_{m,0}-\rho_{m,0}}{\rho_{m,0}}=\frac{1}{k^3}-1
\label{27a19}
\end{equation}
For voids $\Delta_c$ is always negative. Thus, our problem reduces to determining $k$. We need to calculate both the integrals in~(\ref{27a18}) and equate them. The first integral may be calculated analytically:
\begin{equation}
t_U=\frac{2}{3H_0}\dfrac{\arcosh\left(1/\sqrt{\Omega_{m,0}}\right)}{\sqrt{1-\Omega_{m,0}}}
\label{27a20}
\end{equation}
The second one may be transformed into (we use equation~(\ref{27a1}) for $H_0$)
\begin{equation}
t_R=\frac{1}{H_0}\int^k_0 \dfrac{\sqrt{x} dx}{\sqrt{x^3 \Omega_{\Lambda,0}+\Omega_{m,0}\left(1+\frac53\aleph x\right)}},
\label{27a21}
\end{equation}
Equating the age of the Universe calculated in two different ways, $t_U=t_R$, we obtain the final equation that relates \enquote{the factor of additional expansion} of the void, $k$, with the amplitude of the initial perturbation, $\aleph$, where $\Omega_{m,0}$ is the only parameter (we take into account that $\Omega_{m,0}+\Omega_{\Lambda,0}=1$):
\begin{equation}
\frac{2}{3}\dfrac{\arcosh\left(1/\sqrt{\Omega_{m,0}}\right)}{\sqrt{1-\Omega_{m,0}}}=\int^k_0 \dfrac{\sqrt{x} dx}{\sqrt{x^3+\Omega_{m,0}\left(1+\frac53\aleph x - x^3\right)}}.
\label{27a22}
\end{equation}
This equation (together with equation~(\ref{27a19})) solves the problem of the void central density in the spherically symmetric case. It does not imply that the final void is in the linear regime; the equation remains valid for arbitrarily strong final perturbations. As it must be, equation~(\ref{27a22}) does not depend on the choice of the initial moment, which we denoted by \enquote{1}: the central density is completely determined by the amplitude of the initial perturbation, $\aleph$, and $\Omega_{m,0}$.

\begin{figure}
\resizebox{\hsize}{!}{\includegraphics[angle=0]{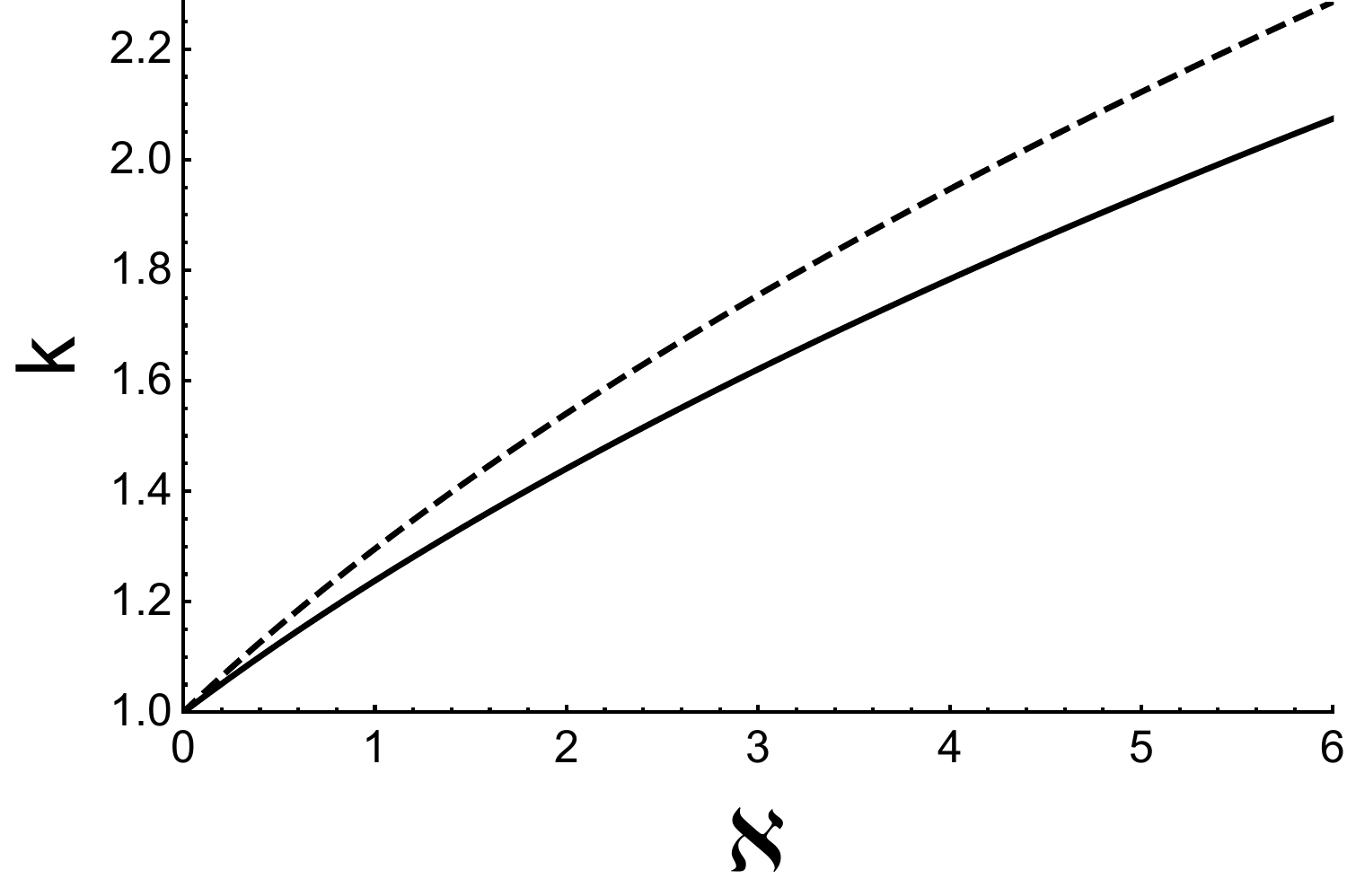}} \caption{The 'factor of additional expansion' of the void, $k$, as a function of the amplitude of initial perturbation, $\aleph$ (eqn.~(\ref{27a5})). The solid line corresponds to our Universe, the dashed one --- to the "no dark energy" case (i.e., to the flat universe with the same value of $H_0$ and $\Omega_{m,0}=1$).}
\label{27fig1}
\end{figure}

\section{The case of an ellipsoidal void}
\label{27section3}
Thus, the central density of a void may be found analytically in the spherically symmetric case. However, this case is too idealized, and it would be more natural to consider he general case of a small initial perturbation of ellipsoidal shape. Will the central density given by equation~(\ref{27a22}) be a reasonable estimate in this case?

First of all, the statement that the outer regions of the void and vast masses of matter surrounding it can hardly create a strong gravitational field at the void center is very likely correct for a general ellipsoidal void as well. Indeed, the density profile of an ellipsoidal void has the form (we take into account that the average density of the Universe is equal to the critical density):
\begin{equation}
\rho=\rho_{c} - \varrho\left(\frac{x^2}{e^2_1}+\frac{y^2}{e^2_2}+\frac{z^2}{e^2_3}\right).
\label{27a23}
\end{equation}
This profile may be divided into a set of thin homoeoids of constant density (a homoeoid is a region bounded by two concentric, similar ellipsoids). A homoeoid homogeneously filled with matter does not create a gravitational field inside itself, and therefore it does not differ from spherical layers in this sense. Thus, if we can neglect the influence of outer layers in the case of a spherical void, we may do the same in the case of an ellipsoidal one.

Second, let us consider how the asymmetry of the distribution~(\ref{27a23}) evolves with time. In the case of an overdensity the answer is well known and can be based on a simple qualitative considerations~\cite{zn2}. Until the perturbation is linear, its shape changes little~\cite{silk1979}. Apparently, this result is valid for an underdensity as well. As an ellipsoidal overdensity reaches the nonlinear regime, its asymmetry rapidly grows. Indeed, in this case the second term in~(\ref{27a23}) is positive. Let us suppose that $e_1$ is the shortest semi-axis in~(\ref{27a23}), i.e., the ellipsoid is compressed the most along the $x$ axis. Then the gravitational acceleration toward the center is also the largest along this axis. The entire velocity field at the nonlinear stage is determined by the gravitational field of the forming structure. Thus, not only the gravitational attraction is the strongest along the shortest axis of the ellipsoid, but also the velocity and acceleration of the contraction are the largest. As a result, the asymmetry rapidly increases, and finally the ellipsoid collapses into a Zeldovich pancake~\cite{zn2}.

\begin{figure}
\resizebox{\hsize}{!}{\includegraphics[angle=0]{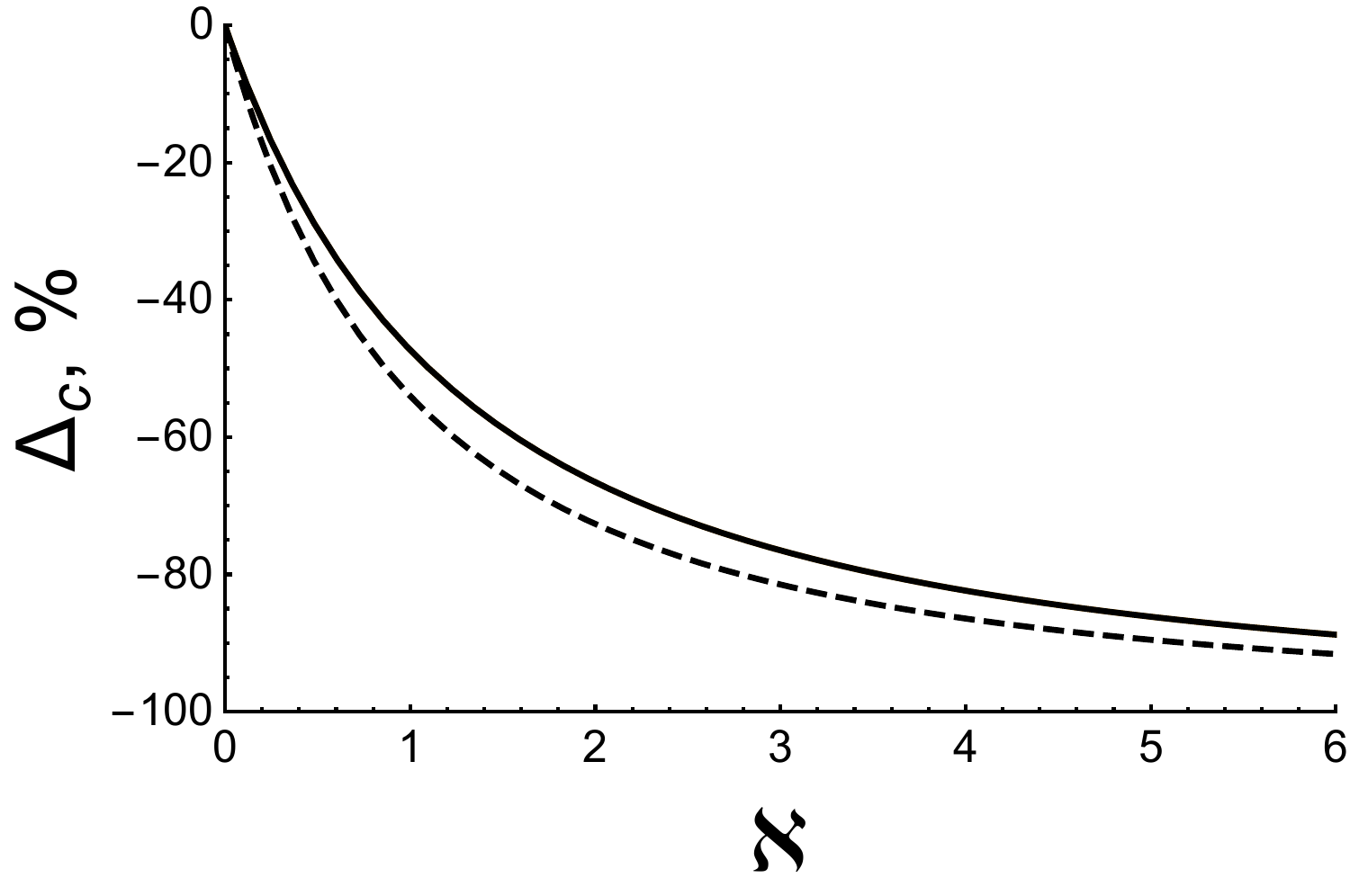}} \caption{The central void underdensity $\Delta_c$ as a function of the amplitude of initial perturbation, $\aleph$ (eqn.~(\ref{27a5})). The solid line corresponds to our Universe, the dashed one --- to the "no dark energy" case (i.e., to the flat universe with the same value of $H_0$ and $\Omega_{m,0}=1$).}
\label{27fig2}
\end{figure}

The same qualitative argument shows that the situation is opposite in the case of a void: the asymmetry at the center decreases with time. Indeed, in this case the second term in~(\ref{27a23}) is negative: the underdensity creates a kind of \enquote{gravitational repulsion} relative to the background Universe. The repulsion is the strongest along the shortest axis of the ellipsoid. As a result, the ellipsoid expands faster in this direction, and the asymmetry decreases with time. Thus, even in the case of an ellipsoidal void its central part tends to spherical symmetry ($e_1=e_2=e_3$). The process of isotropization is relatively slow when the density contrast of the void is small, but accelerates on the nonlinear stage of the void formation.

Third, let us consider the evolution of the central density for an ellipsoidal void. The evolution is determined by the continuity equation:
\begin{equation}
\frac{\dot \rho}{\rho}= - \divi \vec v.
\label{27a24}
\end{equation}
Thus, the density evolution is completely determined by $\divi \vec v$, and we need to find how it depends on the ellipsoid parameters. The density given by~(\ref{27a23}) at the void center depends only on time, since this is an extremum point. The gravitational potential at the void center can be written as~\cite{silk1979}: 
\begin{equation}
\phi=\phi_{u} + \frac12\left(\phi_{xx} x^2+\phi_{yy} y^2+\phi_{zz} z^2\right),
\label{27a25}
\end{equation}
where $\phi_{u}$ is the gravitational potential of the unperturbed Universe. Evidently, $\divi \vec v$ determined by $\phi_{u}$ coincides with that in the spherically symmetric case. Now we need to find $\vec v$ produced by the second term in~(\ref{27a25}), which is actually the perturbation of the velocity field created by the void. The coefficients $\phi_{xx}, \phi_{yy}, \phi_{zz}$ are determined by the central density and the ratios of $e_1$, $e_2$, and $e_3$~\cite{silk1979} and depend only on time. It follows from $\Delta \phi = 4\pi G \rho$ that 
\begin{equation}
\phi_{xx} +\phi_{yy} +\phi_{zz} = 4\pi G \varrho(0)= 4\pi G (\rho-\rho_{c}). 
\label{27a26}
\end{equation}
Thus, though each coefficient $\phi_{ii}$ depends on the void asymmetry, their sum depends only on the central density. 

At the stage when the perturbation is linear, we may calculate the velocity field induced by the perturbation. For instance, for the $x$ component, $dv_x/dt=x\phi_{xx}(t)$. At some previous moment $\tau$
\begin{equation}
\frac{dv_x}{d\tau}=x\frac{a(\tau)}{a(t)}\phi_{xx}(\tau). 
\label{27a27}
\end{equation}
Here we neglect the difference between the scale factor inside and outside the void, which is a reasonable approximation at the linear stage: the particle acquires some additional velocity in the perturbation, but the time is too short to displace it significantly with respect to the background Universe. This approximation fails only when the perturbation becomes nonlinear~\cite{silk1979}. From~(\ref{27a25}) we have:
\begin{equation}
v_i=i\int_0^t \frac{a(\tau)}{a(t)}\phi_{ii}(\tau) d\tau,
\label{27a28}
\end{equation}
where $i=x,y,z$.  We can calculate the velocity divergence:
\begin{equation}
 \frac{dv_x}{dx}+\frac{dv_y}{dy}+\frac{dv_z}{dz}=\int_0^t \frac{a(\tau)}{a(t)}(\phi_{xx}+\phi_{yy}+\phi_{zz})d\tau,\label{27a29}. 
\end{equation}
With the help of equation~(\ref{27a26}) we finally obtain:
\begin{equation}
\divi\vec v=4\pi G\int_0^t \frac{a(\tau)}{a(t)}(\rho(\tau)-\rho_{c}(\tau))d\tau.
\label{27a30}   
\end{equation}
As follows from equation~(\ref{27a28}), the velocity distribution is anisotropic in the case of an ellipsoidal perturbation: the coefficients $\phi_{xx}, \phi_{yy}, \phi_{zz}$ are substantially different. The velocity components are not completely independent, however: they are implicitly related through~(\ref{27a26}), and as a result the velocity divergence~(\ref{27a30}) does not depend on the void ellipticity at all. Thus, equation~(\ref{27a24}) shows that the evolution of the central density of an ellipsoidal void is exactly the same as in the spherically symmetric case, at least at the linear stage of void formation.

To summarize this section: the statement that the outer regions of the void can hardly create a strong gravitational field at the void center is very likely correct for a general ellipsoidal void as well. At the linear stage the evolution of the central density of an ellipsoidal void coincides with that of the spherically symmetric one with the same initial central density. The nonlinear stage requires further study, but the ellipticity rapidly decreases at this stage, making the density profile spherically symmetric. Thus, equations~(\ref{27a19}, \ref{27a22}) for the central density in the spherically symmetric case give at least a very good estimate of the central density even for an ellipsoidal void. 

\section{Are the voids empty?}
\label{27section4}
Figure~\ref{27fig1} presents the relationship~(\ref{27a22}) between the perturbation amplitude $\aleph$ and the \enquote{factor of additional expansion} of the void, $k$; figure~\ref{27fig2} --- the dependence of the  central underdensity, $\Delta_c$, on $\aleph$. The solid lines correspond to the real Universe with $\Omega_{m,0}=0.315$. The dashed lines correspond to the case when the universe has the same $H_0$ as the real Universe and no curvature term, but contains no dark energy (i.e., $\Omega_{m,0}=1$). As one can see, the presence of dark energy suppresses void formation: the voids in the hypothetical universe without dark energy expand faster and have a lower central density.

This is not surprising: the presence of dark energy suppresses the growth of cosmological perturbations, both overdensities and underdensities. This effect may be taken into account by multiplying the perturbation amplitudes by the growth suppression factor $g(z)$ (see, e.g.,~\cite[section 4.4]{gorbrub2} for details). At the matter-dominated stage $g(z)=1$, but even at present, when the dark energy dominates ($\Omega_{\Lambda,0}=0.685$), $g(0)\equiv g_0\simeq 0.788$, i.e., the influence of dark energy on structure formation is moderate.

As equation~(\ref{27a19}) shows, the deeper the initial perturbation is, the larger $k$ is, i.e., the deepest regions of the initial underdensity corresponding to the future void expand the most. This leads to the characteristic void density profile: it is very flat in the center (i.e., the matter density is almost constant in the central part of the void), but grows rapidly toward the edge of the void. For instance, N-body simulations suggest the following profile \cite[eqn. 15]{modelvoid}:
\begin{equation}
\frac{\rho(r)}{\rho_{M,0}}=A_0+A_3\left(\frac{r}{R_V}\right)^3.
 \label{27a31}
\end{equation} 
The authors obtained the best fit values $A_0=0.13\pm 0.01$ and $A_3=0.70\pm 0.03$ for voids with $R_V=8 h^{-1}$~{Mpc}. In reality the central part may be even flatter~\cite{26}. Equations~(\ref{27a19}, \ref{27a22}) define a one-to-one correspondence between the amplitude $\aleph$ of the perturbation that later forms the void and the void's central underdensity $\Delta_c$. One should distinguish $\Delta_c$ from the average void's underdensity $\Delta_{av}$, which is also common in the literature. Obviously, $|\Delta_{c}| \ge |\Delta_{av}|$. Moreover, $\Delta_{av}$ is much worse defined: it strongly depends on how the void boundary is determined. The density profile is very steep near the boundary, and if we slightly shift the boundary (by changing the void criterion), $\Delta_{av}$ changes significantly, while $\Delta_c$ remains the same.

\begin{figure}
\resizebox{\hsize}{!}{\includegraphics[angle=0]{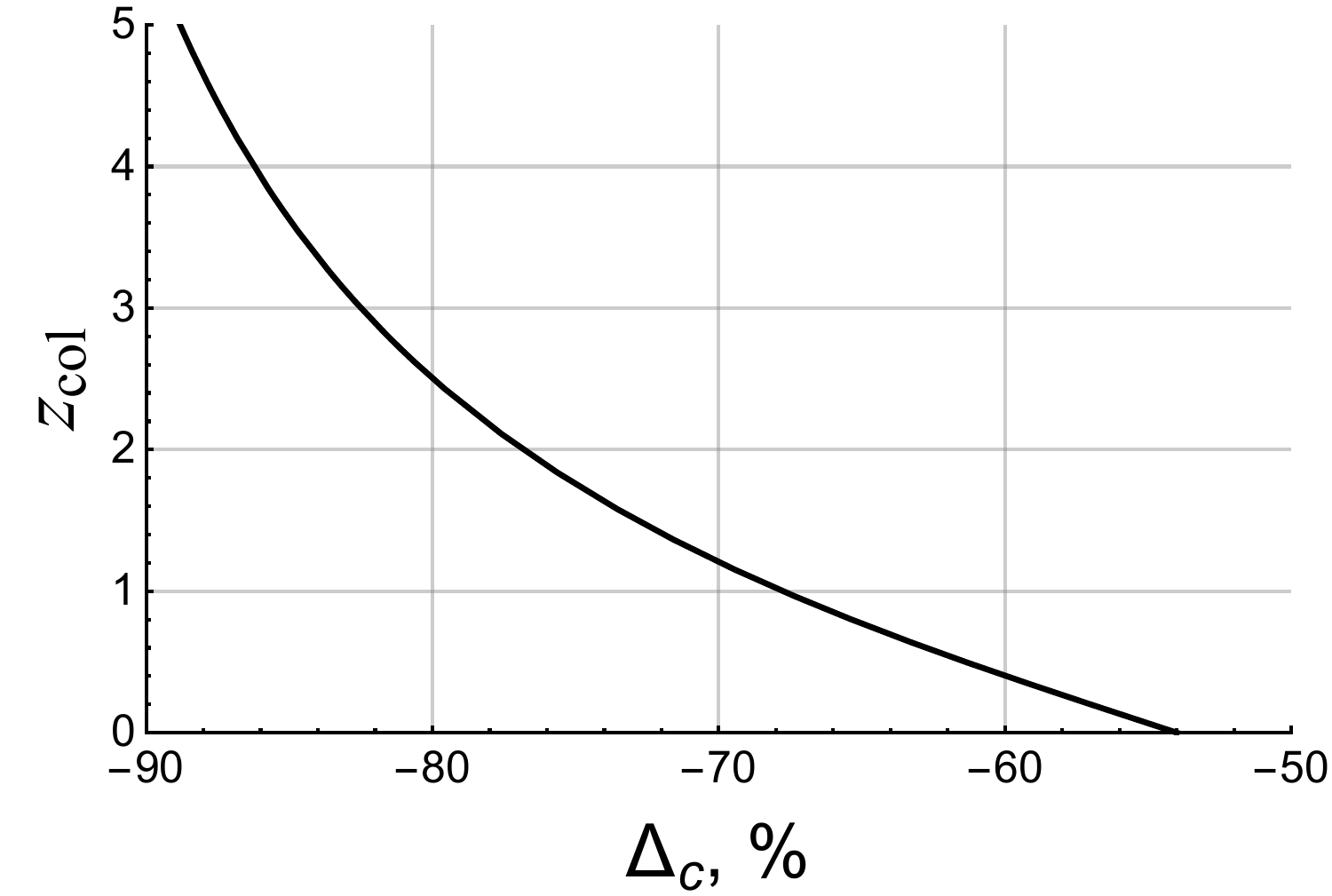}} \caption{The relationship between the central underdensity~(\ref{27a19}) of a void and the formation moment $z_{col}$ of a cluster, assuming that the perturbations forming the void and the cluster had the same size and amplitude $|\aleph|$ when they were linear.}
\label{27fig3}
\end{figure}

In order to compare our results with observations, we need to estimate the values of $\aleph$ corresponding to voids. Following observational results~\cite{pdg2022}, we assume Gaussianity of the primordial perturbations. In particular, this implies that at the linear stage perturbations with the same $|\aleph|$ and size are equally probable, regardless of whether they are under or overdensities. Let us consider an overdensity and an underdensity of the same size and $|\aleph|$ at the linear stage. The underdensity now forms a void of radius $R_0$ and central underdensity $\Delta_c$, which is related to $|\aleph|$ by equations~(\ref{27a19}, \ref{27a22}). The overdensity collapses into a galaxy cluster of mass
\begin{equation}
M_0\simeq \frac43\pi R^3_0 \rho_{m,0}
\label{27a32}   
\end{equation}
at $z=z_{col}$, when its amplitude $\Delta\rho_m/\rho_m\simeq 1$, i.e., 
\begin{equation}
g(z_{col})\dfrac{|\aleph|}{z_{col}+1}=1.
\label{27a33}   
\end{equation}
For example, a galaxy cluster of mass $2\cdot 10^{14} M_\odot$ approximately corresponds to a void of radius $R_0\simeq 11$~{Mpc}. There is a one-to-one correspondence between the central underdensity of the void, $\Delta_c$, and the amplitude $|\aleph|$ of the initial perturbation. Equation~(\ref{27a33}) relates $|\aleph|$ to $z_{col}$ of the cluster. Since we assume that the void and the cluster had the same amplitude of the initial perturbation $|\aleph|$, $\Delta_c$ and $z_{col}$ are also uniquely related, which is shown in figure~\ref{27fig3}.

Thus, if we consider voids and galaxy clusters formed from primordial perturbations of the same size (i.e., the void radius and the cluster mass are related by~(\ref{27a32})) and the same $|\aleph|$ (i.e., their $\Delta_c$ and $z_{col}$ are related as shown in Fig.~\ref{27fig3}), their number densities in the Universe should be equal due to the Gaussianity of the perturbations. Therefore, instead of measuring $\Delta_c$ of the voids of some given radius $R_0$ (which is a difficult task) we may observe $z_{col}$ of the corresponding galaxy clusters. 

Observations \cite[see Fig. 1]{observations2019} show no clusters with a present-day mass exceeding $3\cdot 10^{14} M_\odot$ formed before $z=1.75$. This means that there are no (or extremely rare) voids of radius $R_0\ge 12.5$~{Mpc} with $|\Delta_c| > 75\%$. The first clusters with lower mass may occur at $z<2.5$ \cite{finoguenov2024}, which corresponds to $\Delta_c = -80\%$. Large voids ($R_0\ge 40$~{Mpc}) correspond to clusters of mass $> 10^{16} M_\odot$, which do not exist, because the perturbations of this mass are still linear. Fig.~\ref{27fig3} shows that the central underdensities of the supervoids satisfy $\Delta_c > -55\%$: they are also still linear.

Let us emphasize that we estimated the maximum possible $|\Delta_c|$ of voids in the previous paragraph. Indeed, only the first clusters with a present-day mass $\sim 3\cdot 10^{14} M_\odot$ appear at $z=1.75$, while most of the clusters of this mass collapse much later. By analogy, there may be only rare (one in a hundred) voids with radius $R_0\ge 12.5$~{Mpc} having $\Delta_c \simeq -75\%$, but most of the voids with this radius have a significantly higher central density. 

In order to estimate not the maximum possible $|\Delta_c|$, but its average values for voids with a given radius, we again use the Gaussianity of the primordial perturbations. Let us randomly choose a comoving sphere of radius $L/((z+1)h)$ (so $L/h$ is the present-day radius of the sphere) deep in the matter-dominated stage, when the perturbations corresponding to voids were linear.  The Gaussianity of the perturbations means that the probability $f$ that the average density contrast $\Delta\rho_m/\rho_{m}$ inside $L$ is equal to $\delta_L$ has the Gaussian form:
\begin{equation}
p(\delta_L)=\frac{1}{\sigma_L \sqrt{2\pi}}\exp\left(-\frac{\delta^2_L}{2\sigma^2_L}\right).
\label{27a34}   
\end{equation}
The value of $\sigma_L$ is proportional to $g(z)/(z+1)$. In view of the foregoing, we assume that the distribution of $\aleph$ has the form:
\begin{equation}
f(\aleph)=\frac{1}{\sigma_8 \sqrt{2\pi}}\exp\left(-\frac{\aleph^2}{2\sigma^2_8}\right),
\label{27a35}   
\end{equation}
where $\sigma_8=0.811$ \cite{pdg2022}. Thus, we take the distribution of $\aleph$ to be the same as that of relatively small perturbations ($R_0=L_0/h=8h^{-1}$~{Mpc}~$\simeq 11$~{Mpc}). This size roughly corresponds to small voids, while the largest voids exceed $R_0=70$~{Mpc} \cite{bootesvoid}. Actually, we suppose that $\sigma_L={\it const}=\sigma_8$. In fact the value of $\sigma$ decreases with the object size, and thus our approach somewhat overestimates $\aleph$ and the emptiness of large voids.

\begin{figure}
\resizebox{\hsize}{!}{\includegraphics[angle=0]{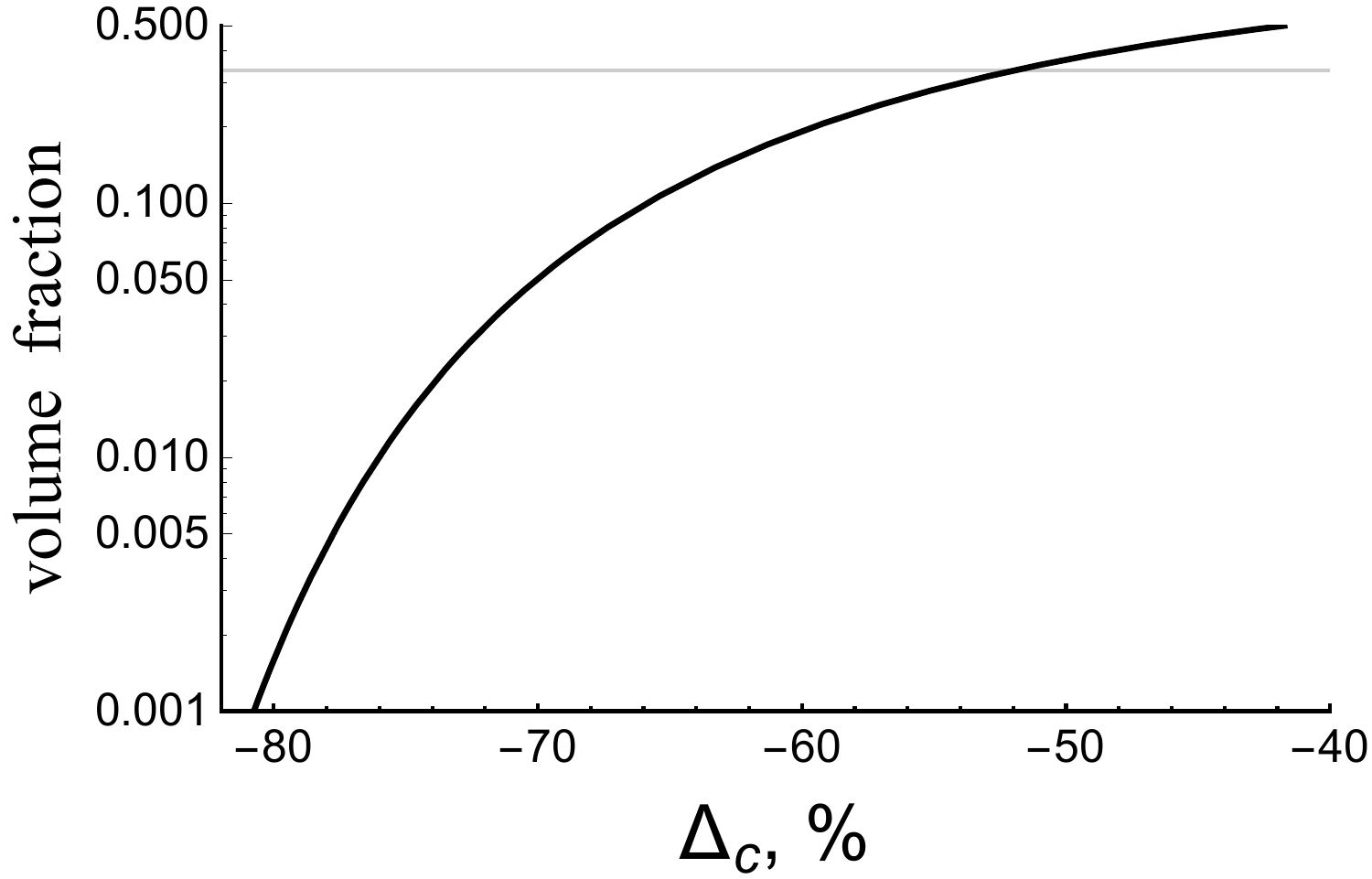}} \caption{The fraction of the Universe volume occupied by regions emptier than $\Delta_c$, as a function of $\Delta_c$. The grey horizontal line corresponds to the volume fraction equal to $1/3$.}
\label{27fig4}
\end{figure}

When the perturbations reach the nonlinear stage, voids expand significantly faster than the Universe on average. As we have discussed, a void has two regions: a large density plateau at the center, where the matter density is almost constant and equal to $q_{m,0}$, and void walls with very steep density profiles. The plateau region expands $k^3(\aleph)$ times as much as the Universe on average. Combining it with~(\ref{27a35}), we find that the volume $V(\Delta_c)$ occupied by the regions with matter underdensity less than $\Delta_c$ is proportional to:
\begin{equation}
V(\Delta_c(\aleph))\propto\frac{k^3(\aleph)}{\sigma_8 \sqrt{2\pi}}\exp\left(-\frac{\aleph^2}{2\sigma^2_8}\right).
\label{27a36}   
\end{equation}
Using~(\ref{27a19}) and~(\ref{27a22}) to express $\aleph$ in terms of $\Delta_c$, we can integrate~(\ref{27a36}) and obtain the fraction of the Universe volume occupied by regions emptier than $\Delta_c$, as a function of $\Delta_c$. This dependence is shown in Fig.~\ref{27fig4}.

It is reasonable to expect that the plateau regions of voids occupy at least $1/3$ of the Universe volume, which corresponds to $\Delta_c \simeq -52\%$. This is likely the most reasonable estimate of the average central underdensity $\Delta_c$ for voids. According to figure~\ref{27fig4}, regions with $\Delta_c \simeq -75\%$ and $\Delta_c \simeq -80\%$ occupy $\sim 1.5\%$ and $\sim 0.1\%$  of the Universe volume, respectively. These fractions are consistent with the above considerations that only one in a hundred small voids may have $\Delta_c \simeq -(75-80)\%$, but such voids are a rare exception, not the rule.

Let us summarize. Most of the voids have $\Delta_c \sim -50\%$; there are no voids with $\Delta_c < -80\%$. For small voids ($R_0\simeq 5-10$~{Mpc}), the underdensity can be significantly deeper than $|\Delta_c| = 50\%$, but the number of voids decreases rapidly as $|\Delta_c|$ increases, and voids with $|\Delta_c| > 80\%$ are practically absent. The largest voids ($R_0 > 40$~{Mpc}) always have $|\Delta_c| < 50\%$: they correspond to overdensities of masses $2\cdot 10^{16} M_\odot$, which exceeds the mass of the richest galaxy clusters. Overdensities of this size are still in the linear regime today, so the corresponding underdensities cannot have entered the nonlinear regime yet. Finally, voids with $R_0= 15-50$~{Mpc} have intermediate parameters: their typical $\Delta_c \sim -50\%$, the number of voids decreases rapidly as $|\Delta_c|$ increases, and the upper limit of $|\Delta_c|$ depends on the void radius (larger voids are less underdense) and is always smaller than $|\Delta_c| =75\%$.

Our estimates differ significantly from the estimates of void densities obtained in N-body simulations. For instance, equation~(\ref{27a31}) is obtained by fitting N-body simulations and suggests an average $\Delta_c = -87\%$ ($A_0=0.13$) even for small voids with $R_0=8h^{-1}$~{Mpc}~$\simeq 11$~{Mpc}. Recent simulations also indicate very underdense voids. \cite{nbody2024} find that voids occupy $\sim 83\%$ of the Universe volume, and their average underdensity is $\Delta_{av} = -65\%$. The fact that voids occupy $83\%$ of the Universe volume in the simulations makes it clear that we are dealing with the average, not the central density of voids. The relationship between $\Delta_{av}$ and $\Delta_c$ depends on the model, but if we use the generally accepted relation~(\ref{27a31}),  $\Delta_{av} = -65\%$ corresponds to $\Delta_c = -90\%$. Thus, simulations consistently give an average of $\Delta_c = -(85-90)\%$ for voids, while this analytical calculation gives an estimate of $\Delta_c = -50\%$ for this value.

%The overestimation of the void emptiness by the N-body simulations is evident from the result of simulations \cite{nbody2024} that the voids with an average underdensity of $\Delta_{av} = -65\%$ occupy $\sim 83\%$ of the Universe volume. Comparing this with equation~(\ref{27a34}), we see that this situation is only possible if $\sigma_L\gg 1$: all the underdense region are already deeply nonlinear ($\Delta_{av} = -65\%$), they have already expanded much more than the Universe on average and occupy $83\%$ of the Universe volume instead of $50\%$ in the linear regime. However, voids are large, and they correspond to $L \gtrsim 8$~{Mpc}. Since $\sigma_L$ decreases with $L$, for voids $\sigma_L \lesssim \sigma_8\simeq 0.811<1$, and the regions with $\Delta_{av} = -65\%$ may occupy only $\sim 10\%$ of the Universe volume, not $83\%$ (see figure~\ref{27fig4}).

%\section{The reason for the discrepancy}

Thus, the analytical estimates obtained in this paper clearly contradict the results of N-body simulations. What is the reason for the discrepancy, and whose estimates are closer to reality? We note that a reliable answer to this question requires a separate, extensive study. Indeed, the model presented in this paper considers a highly simplified situation of an isolated, spherically symmetric void; nevertheless, the analytical approach allows us to find the exact solution. N-body simulations suffer from several numerical artifacts~\cite{vanderbosch2018, 21, 18, 17, 13}, but they can handle realistic voids and their astrophysical environments. Here we briefly outline some arguments suggesting that the analytical estimates~(\ref{27a19}, \ref{27a22}) are closer to reality. Clearly, these considerations are no substitute for a full-fledged study.

The simulations presented in~\cite{nbody2024} show that the voids with an average underdensity of $\Delta_{av} = -65\%$ occupy $\sim 83\%$ of the Universe volume. Comparing this with equation~(\ref{27a34}), we see that this situation is only possible if $\sigma_L\gg 1$: all the underdense region are already deeply nonlinear ($\Delta_{av} = -65\%$), they have already expanded much more than the Universe on average and occupy $83\%$ of the Universe volume instead of $50\%$ in the linear regime. However, voids are large, and they correspond to $L \gtrsim 8$~{Mpc}. Since $\sigma_L$ decreases with $L$, for voids $\sigma_L \lesssim \sigma_8\simeq 0.811<1$, and the regions with $\Delta_{av} = -65\%$ may occupy only $\sim 10\%$ of the Universe volume, not $83\%$ (see figure~\ref{27fig4}).

The results of simulations in~\cite{30} may serve as indirect confirmation of the estimate of the central density of voids obtained in this work. In~\cite{30}, void formation was modeled using semi-analytical methods, and the resulting estimates of the central density confirm the estimates made in the present paper. Moreover, the independence of the central density from the ellipticity of the void is also confirmed in~\cite{30}. However, the study also considered an isolated void and, consequently, neglected tidal effects.

On the other hand, our analytical calculations are also based on several simplifying assumptions, as described above. However, their applicability can be verified much more easily.  Section~\ref{27section3} shows that void ellipticity most likely does not affect the central density of the void. We also show that the tidal influence at the void center is moderate. Moreover, tidal perturbations produced by an external body (at least to the first approximation) deform a dust cloud on which they act without changing its volume. 

However, such considerations are certainly not rigorous and cannot replace careful research. Yet the mere fact that there are undeniable discrepancies between the N-body simulation results and the analytical calculations appears very interesting.

\section{Observational properties of voids}

\begin{figure}
\resizebox{\hsize}{!}{\includegraphics[angle=0]{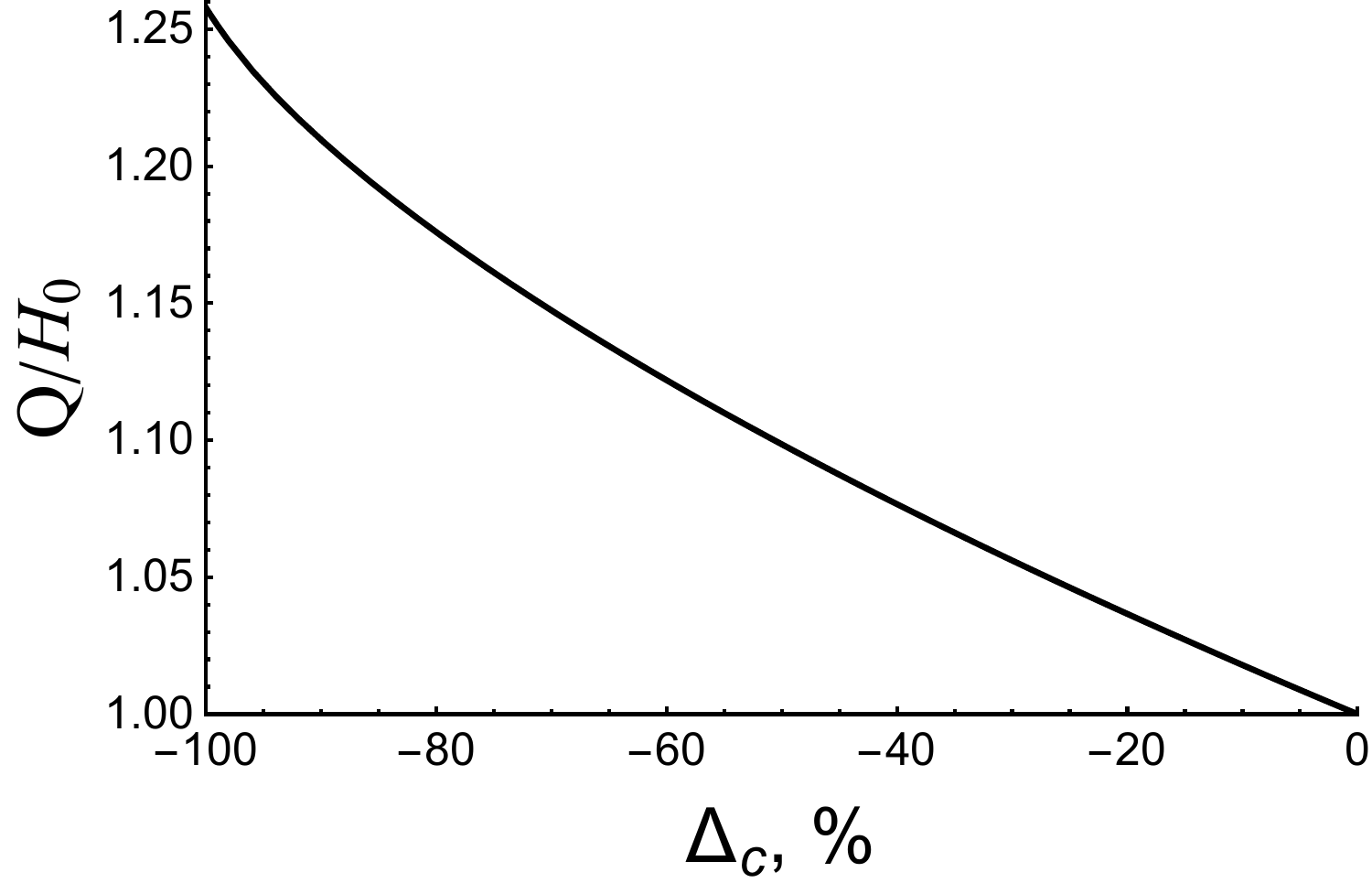}} \caption{The ratio of the Hubble constants at the void center $Q_0$ and of the Universe $H_0$ as a function of the void underdensity $\Delta_c$.}
\label{27fig5}
\end{figure}

As equation~(\ref{27a17}) shows, the \enquote{Hubble constant} $Q$ inside a void differs from that of the Universe, $H$, and depends on $|\aleph|$. It is interesting to check whether we can determine the void underdensity by measuring $Q$. Substituting $k=b_0/a_0$ into~(\ref{27a17}), we obtain the relationship connecting the present-day ratio $Q_0/H_0$ and $\aleph$. Using equations~(\ref{27a19}, \ref{27a22}) we may express $Q_0/H_0$ in terms of $\Delta_c$; the dependence is shown in figure~\ref{27fig5}. As one can see, the \enquote{Hubble constant} inside voids indeed always exceeds $H$. However, the prospects for measuring $\Delta_c$ by observing $Q_0$ are not promising: as $\Delta_c$ changes from $0$ to $-100\%$, $Q_0$ increases only by $\sim 25\%$. Thus, $Q_0$ is not very sensitive to $\Delta_c$, and it is difficult to extract the value of $\Delta_c$ from observations of $Q_0$ inside the void.

Finally, a question arises: if voids contain so much matter, why do they look so empty? Indeed, the density of bright galaxies in voids is tens of times lower than on average in the Universe, while the density of matter in most voids, as we have seen, is only two to three times lower than the cosmic average. The reason for this discrepancy is probably quite simple. We saw that the interior of the void can be considered as a part of a Friedmann space, but its parameters differ from those of our Universe. Firstly, the epoch of matter–dark energy equality occurs somewhat earlier, and so the value of $g_0$ is noticeably lower in it. For this reason alone, the growth of cosmological perturbations in the void is suppressed significantly more strongly than in the Universe on average. Secondly, the \enquote{universe} inside the void corresponds to the open Friedmann model: the three-dimensional space there has Lobachevsky geometry. The open universe expands faster than the flat one, and this also slows down the growth of perturbations.

The problem of the reduced growth of perturbations inside voids certainly requires additional study. Note, however, that both above-mentioned suppression mechanisms are activated comparatively late, when the amplitude of the void-forming underdensity becomes noticeable ($z<5$). Thus, bright structures in voids, such as large galaxies and galaxy clusters, are strongly suppressed. But low-mass structures, such as dwarf galaxies or dark matter clumps, collapse at $z>10$ and therefore form in void in almost the same way as in the rest of the Universe: their present-day number density there is only $k^3\sim 2-3$ times lower than the average in the Universe: it follows the matter density. Thus, voids contain a significant amount of dark and baryonic matter, but it forms only low-mass, faintly luminous objects and a massive diffuse component. Therefore, this matter remains invisible.

Direct astronomical observations would be the best way to verify the above statements. For example, stacked weak-lensing measurements of voids~\cite{2015MNRAS.454.3357C} could directly constrain the total matter density profile independently of luminosity. Nowadays, the observational capabilities of astronomers are rapidly increasing, and deep optical/near-infrared surveys, HI 21-cm observations, and searches for low-surface-brightness or ultra-diffuse galaxies in voids could test whether a substantial population of faint objects is present. Calculating the density of dwarf galaxies in voids could be especially informative. The lower a galaxy's mass, the earlier it forms; consequently, dwarf galaxies originated during an epoch when the void was still deeply in the linear regime of its evolution. Thus, at the time of their formation, the density of dwarf galaxies within the future void was virtually identical to that of the Universe as a whole. Subsequently, as the void formed, the density of dwarf galaxies declined, remaining proportional to the matter density within the void. Therefore, if the void density estimate presented in this article is correct, the density of dwarf galaxies in the void should be only two to three times lower than the cosmic average. If, however, the estimates derived from N-body simulations are accurate, this density should be an order of magnitude lower. Recent observations of the gravitational lensing ~\cite{Martin2026} confirm the analytical estimate obtained in this paper.

We would like to thank the Academy of Finland mobility grant 1341541 for the financial support of this work.

%\bibliographystyle{apsrev4-2}
%\bibliography{article27}

\end{document}